# Magnetic properties of ferromagnetic $Pu_2Pt_3Si_5$

J.-C. Griveau[1,*], E. Colineau[1], D. Bouëxiere[1], K. Gofryk[1,2], T. Klimczuk[1,3] and J. Rebizant[1]

[1]European Commission, Joint Research Centre, Institute for Transuranium Elements; Postfach 2340, 76125 Karlsruhe, Germany

[2]Los Alamos National Laboratory, Los Alamos, New Mexico 87545, USA

[3]Faculty of Applied Physics and Mathematics, Gdansk University of Technology, Narutowicza 11/12, 80-952 Gdansk, Poland

*Abstract*

*The structural, magnetic, and thermodynamic properties of new plutonium based compound, $Pu_2Pt_3Si_5$, are reported. Single crystals produced by a Sn-flux technique have been analyzed showing a ferromagnetic behavior at 58 K. $Pu_2Pt_3Si_5$ crystallizes in the $U_2Co_3Si_5$-type orthorhombic Iabm structure (72) with atomic parameters a= 9.9226(2) Å, b= 11.4436(2) Å and c= 6.0148(1) Å. The effective ($\mu_{eff}$~ 0.74 $\mu_B$) and saturated ($\sigma_{sat}$~ 0.32 $\mu_B$/Pu) moments as well as the linear Sommerfeld coefficient ($\gamma_e$~ 2 $mJ.mol^{-1}.K^{-2}$/Pu) could point towards 5f localization in this material.*

Keywords: actinide alloys and compounds, magnetically ordered materials, heat capacity, phase diagrams, phase transitions

[*]Corresponding author:
Jean-Christophe Griveau: jean-christophe.griveau@ec.europa.eu
Tel. +49(0)7247 951 428
Fax:+49(0)7247 951 99 428

## 1. INTRODUCTION

One can observe an analogy between rare earths (RE) and actinides based compounds, especially in the case of intermetallics containing transition metals, where the localization and delocalization features are taking place [1-2]. The well known and extensively examined superconducting Ce- and Pu-"115" families are good examples of these phenomena, where the superconductivity and magnetism are intrinsically linked [3-7]. Therefore, exploring the basic properties of a class of materials by substituting the *4f* (RE) atom by a *5f* (Actinides) atom is an interesting way to understand the impact of hybridization between *f* and *d* electrons on the ground properties. An interesting family, where comparison between RE and Actinide systems can be done, considering localization and itinerancy, is the $R_2M_3X_5$ family (R=Rare Earth or Actinide, T=transition metals and X=Si or Ge). In several $Ce_2M_3X_5$ systems (M =Ni, Rh, Ir, Pt, Pd; X= Si, Ge), due to the competition between Kondo effect and Ruderman–Kittel–Kasuya–Yosida (RKKY) interactions, a wide variety of behaviors can be observed [8-

10]. For instance, mixed valence behavior has been reported for $Ce_2Ni_3Si_5$ [11], $Ce_2Rh_3Si_5$ [12], and $Ce_2Ir_3Si_5$ [13] while reentrant superconductivity below $T_N$ occurs in the antiferromagnet $Tm_2Fe_3Si_5$ [14]. $Lu_2Ir_3Si_5$ also becomes superconductoring below 3.5 K [15] and shows unusual thermodynamic and structural behaviors, such as a pronounced structural phase transition that develop 140 and 200 K [16]. $Er_2Ir_3Si_5$ exhibits Charge Density Wave (CDW) transition at 150 K [17]. In addition, heavy fermion behavior is observed in $Yb_2Fe_3Si_5$ [18], and $Pr_2Rh_3Ge_5$ [19], whereas $Ce_2Rh_3Ge_5$ and $Ce_2Ir_3Ge_5$ are antiferromagnetically ordered dense Kondo systems [20, 21]. Considering the actinide-based materials only, few systems are reported and even less, essentially Th and U based systems [22-24], have been examined extensively. For instance, only $U_2Rh_3Si_5$ has been produced in a single crystal form what was crucial to understand the unusual character of the magnetic phase transition at 25.6 K [25] which appears to be of first order and strongly coupled to the lattice [26]. In the case of transuranium based compounds even fewer systems are reported and essentially only their crystallographic parameters are known [22].

The $R_2M_3X_5$ family crystallizes in three types of structures. The Ce and Yb based compounds can mainly be found in an orthorhombic structure of $U_2Co_3Si_5$-type with space group Ibam that may be described as an intergrowth of $CaBe_2Ge_2$ and $BaNiSn_3$ crystal structure. The uranium based compounds may also present a slightly distorted monoclinic structure - called $Lu_2Co_3Si_5$ variant with space group C2/c. Finally, $Pu_2T_3Si_5$ (T=Fe, Tc, Ru) [27] and $Np_2Re_3Si_5$ [28], the only Transuranium representatives of the $An_2T_3X_5$ family (mainly silicides), crystallise all in a tetragonal $Sc_2Fe_3Si_5$-type with space group P4/mnc. It is interesting to note that the local coordination of *4f/5f* atoms for all these systems is similar. Recently, two new plutonium silicides have been reported, $Pu_2Ni_3Si_5$ and $Pu_2Co_3Si_5$ [29]. They crystallize in slightly different crystallographic structures, $U_2Co_3Si_5$- (space group Ibam) and monoclinic $Lu_2Co_3Si_5$-type (space group C2/c), respectively. Both present relatively complex magnetic ordering and enhanced Sommerfeld coefficients $\gamma_e$ close to 100 $mJ.mol^{-1}.K^{-2}$/Pu atom. This may be compared to the uranium compounds of the same family, where non-magnetic moderate heavy fermion behavior has been reported. In addition, it seems that the transition metal element has an impact on the crystallographic structure, since these new Pu compounds did not show the $Sc_2Fe_3Si_5$ type of structure. It would be interesting to examine other Pu-based systems to determine some systematics of the influence of the transition metal. Therefore, we initiated a study on a new Pu-based silicide system from the $An_2T_3X_5$ family. Here, we report on the crystallographic and low temperature magnetic and thermal properties of a new member this family, namely $Pu_2Pt_3Si_5$.

## 2. EXPERIMENTAL DETAILS

The starting materials, plutonium metal with 3N purity, platinum with 5N purity and silicon with 9 N purity, were placed in an alumina crucible together with tin metal (Sn) with the atomic ratio Np : Pt : Si : Sn - 1 : 4 : 20 : 40. The crucible was then encapsulated in a quartz ampoule under high vacuum and install vertically in an electrical furnace. The temperature was slowly increased to 1050°C then decreased down to 750°C at a low cooling rate of 1°C/h. At this temperature the furnace was switched off. The Sn-flux was first removed by warming up pieces extracted from the crucible together with glass wool at 200°C in a metallic container. Remaining ingots were washed into hydrochloric acid to remove the remaining Sn-flux. The so-obtained material was examined by powder x-ray diffraction by using a Bruker D8 diffractometer with the monochromated Cu $K_{\alpha 1}$ radiation ($\lambda$ = 1.540 59 Å), equipped with a Vantec detector. The powder diffraction pattern (Fig. 1) was analyzed by Rietveld profile refinement method [30] using HighScoreExpert software. The result indicates that three phases are present at ~ mm$^3$ scale. The main crystalline phases are Si (37%), PtSi (33%), and $Pu_2Pt_3Si_5$ (30%). Complementary phase composition analysis (EDX) was performed by energy dispersive x-ray analysis on a Philips XL40 scanning electron microscope (SEM). The microprobe analysis indicates the presence of tiny single crystals ~ 100 x 100 x 100 μm$^3$ size of $Pu_2Pt_3Si_5$ of good homogeneity embedded in blocks of PtSi and Si in the same percentage ratio as obtained by XRD powder diffraction. It is important to note that both, Si and PtSi, are diamagnets with $\chi_{Si}$ = -3.12x10$^{-6}$ emu.mol$^{-1}$ and $\chi_{PtSi}$ = -63x10$^{-6}$ emu.mol$^{-1}$, respectively [31, 32] and that their magnetic contribution are negligible (<1%) to the global signal during magnetization measurements (see below).

The magnetic properties were determined using a Quantum Design (QD) MPMS-7 device in the temperature range 2–300 K and in magnetic fields up to 7 T. The specific heat experiments were performed by the relaxation method in a QD PPMS-9, within the temperature range 3.5–300 K and in magnetic fields up to 9 T. Due to the contamination risk generated by the radio toxicity of plutonium, all operations of preparation and encapsulation have been performed in glove boxes under an inert $N_2$ atmosphere.

## 3. RESULTS

### a) Crystal structure

Small single crystals of $Pu_2Pt_3Si_5$ of typical size of 50 × 50 × 20 μm$^3$ were extracted and examined on Enraf-Nonius CAD-4 diffractometer with the graphite monochromatized Mo K$\alpha$ radiation. The crystal structure was solved by the direct method using the SHELX-97 [33]. The results indicate that the new phase, $Pu_2Pt_3Si_5$, crystallizes with the orthorhombic Iabm structure (72) with atomic parameters: a= 9.9226(2) Å, b= 11.4436(2) Å and c= 6.0148(1) Å. The crystallographic parameters, atomic positions, and structural information are given in Tables 1 and 2.

### b) Magnetization

Fig. 2 presents the magnetization curves of $Pu_2Pt_3Si_5$ measured in magnetic field of 10 and 70 kOe. The temperature behavior of the magnetization indicates that $Pu_2Pt_3Si_5$ orders ferromagnetically below $T_C$ ~ 60 K. Above 100 K the inverse molar magnetic susceptibility, H/M ~ 1/$\chi$ determined at H= 70 kOe, can be well fitted by a modified Curie–Weiss law [34] $1/\chi = 1/(\chi_0 + C/(T-\theta_p))$ (see the inset in Fig. 2), yielding an effective magnetic moment $\mu_{eff}$ = 0.78 $\mu_B$, paramagnetic Curie temperature $\theta_p$= 53.5 K, and $\chi_o$=0.00322 emu.mol$^{-1}$. The latter term usually includes the core-electron diamagnetism, the Pauli paramagnetism, and Van Vleck contribution. The positive value of the paramagnetic Curie temperature is consistent with ferromagnetic ordering in this material. The value of $\mu_{eff}$ = 0.78 $\mu_B$ is close to the Hund's rule value for $Pu^{3+}$ (0.84 $\mu_B$) and suggest rather localised *5f*-electrons in this compound. The difference in the magnetization obtained at zero-field-cooled (ZFC) and field-cooled (FC) modes can be explained by the progressive alignments of domains when magnetic field is applied to the sample. In this material, with probable strong anisotropic features (orthorhombic structure), one should expect a strong magneto anisotropy with domains settled along the main easy directions. Fig. 3, presents hysteresis loops, obtained at different temperature in the ordered state. The estimation of the saturated magnetic moment, by taking magnetization at 2 K and 70 kOe, gives $\sigma_{sat}$= 0.35 $\mu_B$/Pu atom. This value is similar to other Pu-based ferromagnetic compounds such as $PuFe_2$ (0.4 $\mu_B$) [34] and PuSb (0.3 $\mu_B$) [35]. We observe an increase of the coercitive field, reaching $H_c$~ 1.9 kOe at 5 K, and an increase of the saturated magnetization when cooling down. Considering the value of the coercitive field and the shape of the magnetization $\sigma(H)$ for the first magnetization curve (see inset Fig. 3), we can suggest that the pinning or lasting immobilization of the magnetic walls, along the main directions of the crystals grains are the principal processes of coercivity in the material [36]. In order to determine the value of the Curie temperature $T_C$, we performed Arrot plots analysis as shown in Fig. 4. We find that $T_C$ is close to 58 K (see dashed line in the figure). The so-obtained $T_C$ is close to the paramagnetic Curie temperature $\theta_P$ as expected for ferromagnetic systems. Inset of Fig. 4 presents the reduced spontaneous magnetization $\sigma_s(T)/\sigma_s(0)$ as a function of reduced temperature $T/T_C$ obtained from the Arrot plots analysis for $Pu_2Pt_3Si_5$. In the same figure we show the theoretical curves obtained either from the 2D [37] or 3D [38] Ising models as well as within the molecular field approximation (MFA) [39]. As shown, the MFA model seems to describe the magnetic properties of $Pu_2Pt_3Si_5$ much better than the other models.

### c) Heat capacity

Fig. 5 presents the temperature dependence of the heat capacity of a single crystal of $Pu_2Pt_3Si_5$ characterized on a CAD-4 diffractometer and identified as single phase. A pronounced anomaly at $T_C$ = 58 K is observed in agreement with the magnetic measurements. At 300 K the heat capacity has a slightly reduced value (~ 210 J.mol$^{-1}$K$^{-1}$) vs. Dulong-Petit limit, i.e., $C_p$ = 3nR = 244 J/mol K, where n = 10 is the number of atoms per formula unit and R is the gas constant, indicating that the Debye temperature is slightly smaller than 300 K. Surprisingly, the low temperature contribution is really small as for metallic systems and the determination of the Sommerfeld contribution ($C_p/T \sim \gamma_e + \beta T^2$) leads to $\gamma_{e/Pu}$ = 2 mJ.mol$^{-1}$K$^{-2}$/ Pu, and $\beta$ = 0.85 mJ.mol$^{-1}$.K$^{-4}$ (see the inset in Fig. 4). We estimate the Debye temperature by the relation $\theta_D \sim (12\pi^4 N_A n k_B/5\beta)^{1/3}$, where n is the number of atoms per formula unit, $N_A$ is Avogadro Number, and $k_B$ is the Boltzman constant. For $Pu_2Pt_3Si_5$, by taking $\beta$ = 0.85 mJ.mol$^{-1}$.K$^{-4}$ this leads to $\theta_D$ ~ 286 K. The so-obtained value of $\theta_D$ is close to the one reported for the Rare Earths counterpart, with the same structure, $Lu_2Ir_3Si_5$ ($\theta_D$ ~ 320 K) [15]. In $Pu_2Pt_3Si_5$ the low temperature part of the electronic specific heat, $\gamma_{e/Pu}$~2 mJ.mol$^{-1}$K$^{-2}$/Pu, is very small when comparing to the other actinide based intermetallics, especially to similar $Pu_2Ni_3Si_5$ (~85 mJ.mol$^{-1}$K$^{-2}$/Pu) and $Pu_2Co_3Si_5$ (~100 mJ.mol$^{-1}$K$^{-2}$/Pu) dense Kondo systems. This suggests well localized *5f*-electrons in $Pu_2Pt_3Si_5$. Fig. 6 shows the temperature dependence of the specific heat of $Pu_2Pt_3Si_5$ in the vicinity of the magnetic transition taken at different magnetic fields. As seen, under magnetic field, the peak on $C_p(T)$ is suppressed and shifted toward higher temperatures as expected for ferromagnetic systems.

## 4. DISCUSSION

When looking at plutonium based systems one can see that not so many compounds show magnetic ordering. Despite the fact that it is much more difficult to examine basic properties of plutonium based compounds than of Th, U and Np based materials, the majority of Pu based compounds, and not only intermetallics, have a tendency to present a non magnetic ground state [40]. One point to consider is that the self-decay and self-irradiation processes are generating defects into the material. This can very fast prevent an access to the magnetic features, especially at low temperature, where ferromagnetic or antiferromagnetic order should take place and could be smeared out by defects induced. For the majority of the reported Pu-based phases only the crystallographic structure is known. The magnetic order and its type are only well established in a few cases, and among them, a very small fraction is ferromagnetic (Table 3). At first glance it appears that ferromagnetism is mainly observed in the Pu binaries [41-49]. One reason for this is a clear lack of knowledge of basic properties at low temperature of Pu-based ternaries [29,50].

The actinides interdistance $d_{An-An}$ has been shown to be an important parameter to observe magnetism in actinides compounds. This has been corroborated by thorough studies in uranium based systems and then extrapolated to neptunium and plutonium systems [51]. Similarly to light actinides, plutonium systems show a similar trend, however several exceptions exist. For the Laves phase $PuFe_2$, the Pu inter-distance is below the Hill limit (3.4 Å). That should not allow magnetic ordering, but a strong *3d-3d* coupling is compensating this aspect and high $T_C$ values (+564 K) is observed essentially due to iron, especially when compared to the others members of the $AnFe_2$ familly [An=U, Np, Pu, Am][46]. Also $PuPt_2$ presents ferromagnetic properties below $T_C = 6$ K despite the fact that its $d_{Pu-Pu}$ spacing is below the Hill limit. To conclude on the main trend in Pu-based ferromagnetic binaries, one can observe a tendency of these compounds to present lower magnetic order when compared to their U and Np counterparts [40]. The last two recently reported Pu-based magnetic ternaries [29], for instance, show an ordering temperature close to 60 K and $Pu_2Pt_3Si_5$ also fits to this trend.

Finally, focusing on the magnetic properties of the $An_2T_3Si_5$ compounds [52-58] (An=Th, U, Np, Pu – T=transition metals), it appears that only few systems present magnetic ordering (see Table 4) among the 16 systems reported. In the case of the uranium based system, despite the fact that $d_{U-U}$ for uranium compounds is always above Hill limit, magnetic order is rarely observed. Surprisingly, and contrarily to the uranium based materials, all of the transuranium ternaries studied at low temperature display a magnetic ordering. Unfortunately, nothing is known for the neptunium $Np_2T_3Si_5$ phases and some $Pu_2T_3Si_5$ but as all the minimum interdistances $d_{An-An}$ between actinide atoms (Np or Pu) are beyond the Hill limit, we can suppose a tendency towards magnetic order, that should be observed at moderated low temperature (~ 50-70 K) as demonstrated for the last three reported plutonium systems $Pu_2Ni_3Si_5$, $Pu_2Co_3Si_5$ and $Pu_2Pt_3Si_5$.

## 5. SUMMARY and CONCLUSION

Single crystals of a new plutonium based intermetallics, $Pu_2Pt_3Si_5$, have been synthesized by

Sn flux and magnetic properties have been examined at low temperature and under magnetic field by magnetization and heat capacity. This compound crystallizes in the $U_2Co_3Si_5$-type orthorhombic Iabm structure (72) with lattice parameters a= 9.9226(2) Å, b= 11.4436(2) Å and c= 6.0148(1) Å. All the measurements obtained indicate that this new phase is a ferromagnet at $T_C$ ~ 58 K. In addition, the value of the effective magnetic moment, being close to the $Pu^{3+}$ value, together with a very small low temperature electronic specific heat could indicate a well localization of 5$f$-electrons in $Pu_2Pt_3Si_5$.

.

**Acknowledgements**
*The high-purity Pu metal required for the fabrication was made available in the framework of a collaboration with the Lawrence Livermore and Los Alamos National Laboratories and the U.S. Department of Energy.*

Table 1
Structure and profile data

| | |
|---|---|
| Formula sum | $Pu_{8.00}Pt_{12.00}Si_{20.00}$ |
| Formula mass/ g/mol | 4854.7900 |
| Density (calculated)/ g/cm$^3$ | 11.8018 |
| F(000) | 1968.0000 |
| Space group (No.) | I b a m (72) |
| Lattice parameters | |
| a/ Å | 9.9226(2) |
| b/ Å | 11.4436(2) |
| c/ Å | 6.0148(1) |
| alpha/ ° | 90 |
| beta/ ° | 90 |
| gamma/ ° | 90 |
| V/ 10$^6$ pm$^3$ | 682.98170 |
| d(Pu-Pu)/ Å | 4.10 |

Table 2

| Atom | Wyck. | x | y | z |
|---|---|---|---|---|
| Pu | 8j | 0.2696(2) | 0.3703(3) | 0.000000 |
| Pt1 | 4b | 0.500000 | 0.000000 | 0.250000 |
| Pt2 | 8j | 0.1123(2) | 0.1366(3) | 0.000000 |
| Si1 | 4a | 0.000000 | 0.000000 | 0.250000 |
| Si2 | 8g | 0.000000 | 0.274(2) | 0.250000 |
| Si3 | 8j | 0.346(2) | 0.119(2) | 0.000000 |

Table 3:

| Compound | T (AF/F) (K) | d(Pu-Pu)(Å) | $\mu_{eff}(\mu_B)$/Pu | $\sigma_{sat}(\mu_B)$/Pu | $\Theta_p$(K) | $\chi_0$ (10$^{-6}$ emu.mol$^{-1}$) | Ref. |
|---|---|---|---|---|---|---|---|
| PuSi | 72(F) | 3.62 | 0.72 | 0.33 | +79 | 357 | [41] |
| PuP | 126(F) | 3.99 | 0.87 | >0.42 | +130 | 197 | [42] |
| PuAs | 123(F) | 4.14 | 0.98 | 0.67 | +129 | 330 | [43] |
| PuSb | 85(AF)/67(F) | 4.41 | 0.86 | 0.65 | +90 | 200 | [44] |
| PuPt | 44(AF)/19(F) | 3.73 | 0.84 | 0.22 | +24 | | [45] |
| PuFe$_2$ | 564(F) | **3.11** | **3.65** | 0.45 | +599 | 270 | [46] |
| PuGe$_2$ | 35(F) | 4.02 | - | >0.14 | - | - | [47] |
| PuPt$_2$ | 6(F) | **3.31** | 0.89 | 0.2 | +6 | 240 | [48] |
| PuGa$_3$$^{(DO19)}$ | 20(F) | 4.13 | 0.77 | 0.21 | +15.6 | 175 | [49] |
| PuAsSe | 126(F) | 4.00 | 0.5 | >0.1 | - | 500 | [50] |
| PuAsTe | 125(F) | 4.27 | 0.64 | >0.1 | - | 400 | [50] |
| Pu$_2$Ni$_3$Si$_5$ | 65(F)/35(AF) | 3.92 | 0.98 | 0.08 | +40 | - | [29] |
| Pu$_2$Pt$_3$Si$_5$ | 58(F) | 4.10 | 0.74 | 0.35 | +53 | 3220 | This work |

Table 4:

| Compound | Structure-type | a (Å) | b (Å) | c (Å) | β (°) | d(An-An) (Å) | V (Å³) | ρ (g.cm³) | $T_{ord}$ (K) LT behaviour | $\gamma_e$ (mJ.mol⁻¹An⁻¹.K⁻²) | $\mu_{eff}$ ($\mu_B$)/Pu | $\theta_p$ (K) | Ref. |
|---|---|---|---|---|---|---|---|---|---|---|---|---|---|
| Th₂Rh₃Si₅ | Lu₂Co₃Si₅ | 11.525(1) | 11.643 | 5.929 | 120.94 | - | 682.4 | 8.89 | | - | | | [52] |
| Th₂Re₃Si₅ | Sc₂Fe₃Si₅ | 10.98(1) | - | 5.758(8) | - | - | 694.2 | 11.13 | | | | | [53] |
| U₂Fe₃Si₅ | Lu₂Co₃Si₅ | 10.848 | 11.476 | 5.5518 | 119.4 | 3.83 | 602.1 | 8.65 | | - | - | - | [54] |
| U₂Co₃Si₅ | U₂Co₃Si₅ | 9.591 | 11.129 | 5.617 | - | 3.85 | 599.5 | 8.79 | SF | - | 3.03 | -97 | [55] |
| U₂Tc₃Si₅ | Sc₂Fe₃Si₅ | 10.856(1) | - | 5.531(1) | - | 3.82 | 651.8 | 9.31 | 11(AF) | - | - | - | [28] |
| U₂Ru₃Si₅ | Lu₂Co₃Si₅ | 11.092 | 11.762 | 5.707 | 119.3 | 3.93 | 649.3 | 9.41 | - | 72 | 3.21 | -208 | [55] |
| U₂Rh₃Si₅ | Lu₂Co₃Si₅ | 11.404 | 11.435 | 5.815 | 120.7 | 3.91 | 652.0 | 9.42 | 25(AF) | 22 | 3.14 | -62 | [55] |
| U₂Re₃Si₅ | Sc₂Fe₃Si₅ | 10.88(1) | - | 5.524(8) | - | 3.99 | 653.9 | 11.94 | - | - | - | - | [56] |
| | Sc₂Fe₃Si₅ | 10.867(7) | - | 5.518(4) | - | - | 651.6 | 11.98 | | - | - | - | [53] |
| U₂Os₃Si₅ | Lu₂Co₃Si₅ | 11.103 | 11.726 | 5.760 | 118.9 | 3.95 | 656.5 | 12.01 | - | - | 2.78 | -223 | [54] |
| Np₂Re₃Si₅ | Sc₂Fe₃Si₅ | 10.89(1) | - | 5.509(8) | - | 3.89 | 653.3 | 11.92 | | - | - | - | [28] |
| Pu₂Fe₃Si₅ | Sc₂Fe₃Si₅ | 10.432(2) | - | 5.514(1) | - | 3.73 | 600.1(3) | 8.70 | - | - | - | - | [27] |
| Pu₂Co₃Si₅ | Lu₂Co₃Si₅ | 10.957(10) | 11.498(11) | 5.560(5) | 119.256(9) | 3.87 | 611.2(10) | 8.708 | 37(AF)/5(AF) | 95 | - | - | [29] |
| Pu₂Ni₃.₁₁₍₁₎Si₄.₈₉₍₁₎ | U₂Co₃Si₅ | 9.583(10) | 11.280(11) | 5.661(6) | - | 3.92 | 612.0(10) | 8.725 | **65(F)/35(AF)** | 85 | 0.98 | 40 | [29] |
| Pu₂Tc₃Si₅ | Sc₂Fe₃Si₅ | 10.917(4) | - | 5.538(2) | - | 3.91 | 651.8 | 9.21 | - | - | - | - | [27] |
| Pu₂Re₃Si₅ | Sc₂Fe₃Si₅ | 10.929(7) | - | 5.535(4) | - | 3.91 | 661.1 | 11.87 | - | - | - | - | [28] |
| | Sc₂Fe₃Si₅ | 10.926(3) | - | 5.534(2) | - | - | 660.7(5) | 11.83 | | - | - | - | [27] |
| Pu₂Pt₃Si₅ | U₂Co₃Si₅ | 9.9226(2) | 11.4436(2) | 6.0148(1) | - | 4.10 | 682.98(7) | 11.80 | **58(F)** | 2 | 0.78 | 53 | This work |

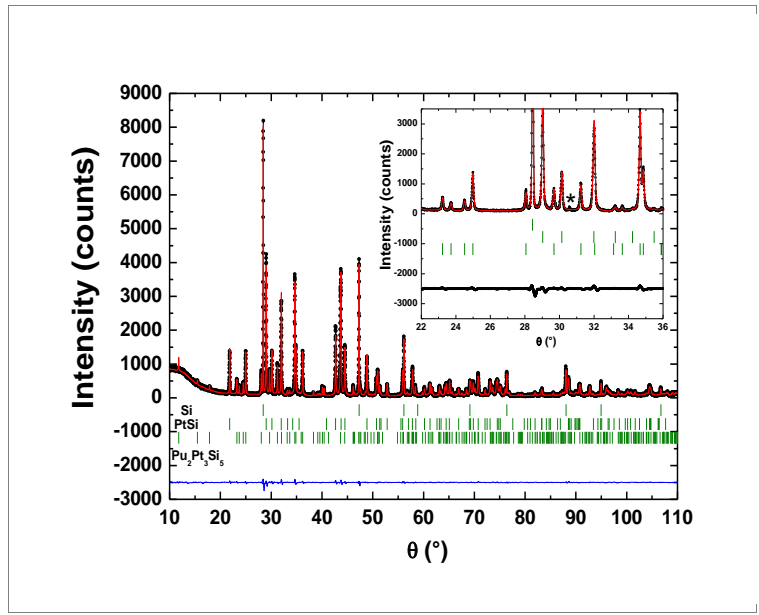

Figure 1: (Color online) X-ray powder diffraction pattern for ingots obtained (see text). The solid red line through the experimental points is the Rietveld refinement profile calculated for mixed phases Si, PtSi and Pu₂Pt₃Si₅. The blue line (bottom) corresponds to the difference between the measured and calculated Rietveld refinement.
Inset: Refinement results obtained around 30° and the presence of an extra peak that cannot be assigned (marked by an asterisk).

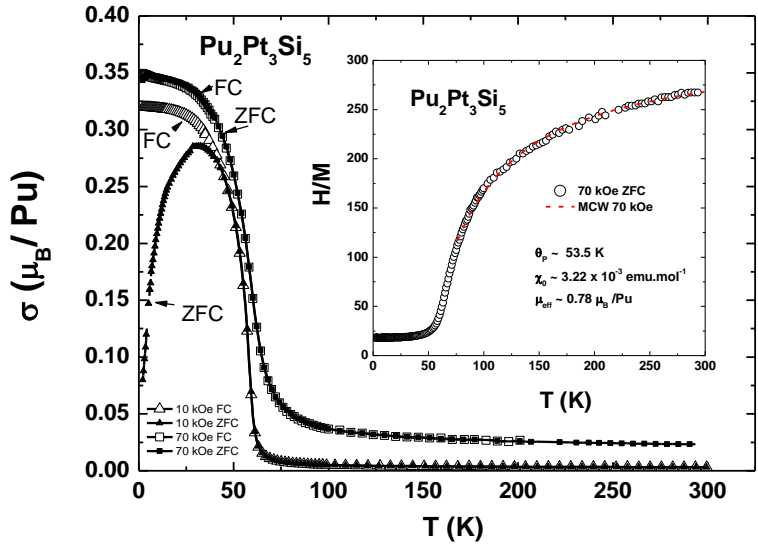

Figure 2: (Color online) The magnetization curves of $Pu_2Pt_3Si_5$ measured in zero-field-cooled (ZFC) and field-cooled (FC) modes for 10 and 70 kOe, respectively.
Inset: the inverse molar susceptibility $1/\chi \sim H/M$. The dashed line is a fit to the modified Curie Weiss law $\chi=\chi_0+C/(T-\theta_p)$ (see text).

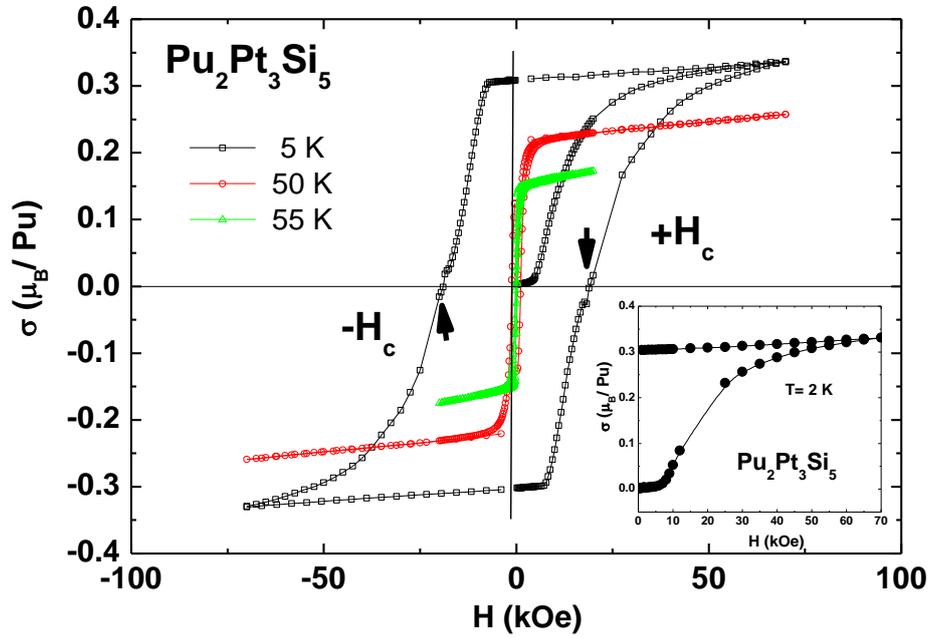

Figure 3: (Color online) Hysteresis loops close to $T_C$ and at very low temperature below magnetic order. We observe the increase of the hysteretic range when cooling down.
Inset: First magnetization curve at 2 K for saturated moment determination.

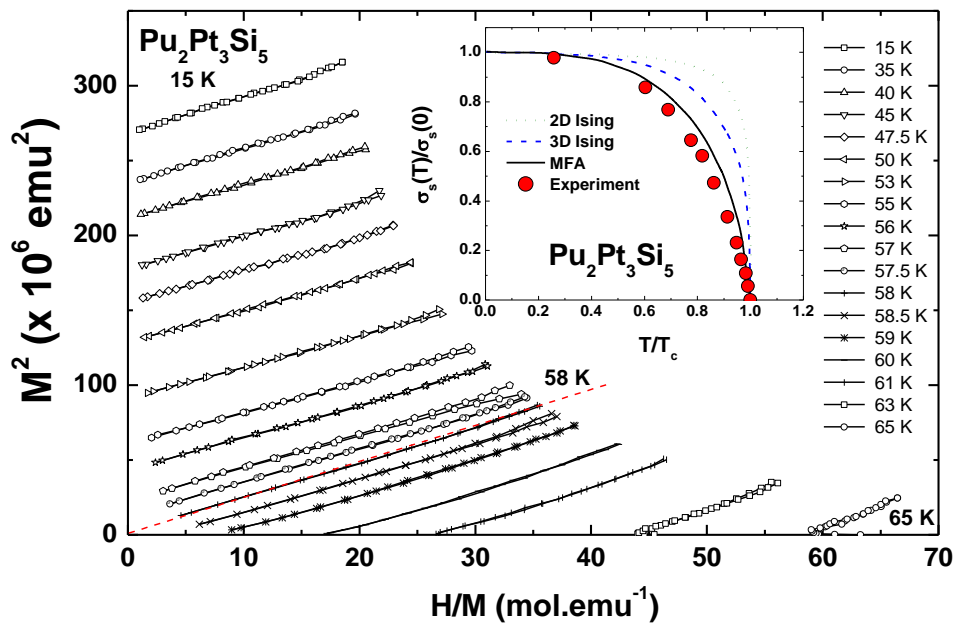

Figure 4: (Color online) Arrot plot for Pu$_2$Pt$_3$Si$_5$. The various isotherms were taken at the temperatures given in the figure. The red dashed line corresponds to the results expected at the Curie temperature.

Inset: reduced magnetization vs. reduced temperature derived from the Arrott plot (solid circles). The dotted, dashed and solid lines represent the theoretical dependences calculated within molecular field approximation 2D, 3D Ising and MFA model respectively (see text).

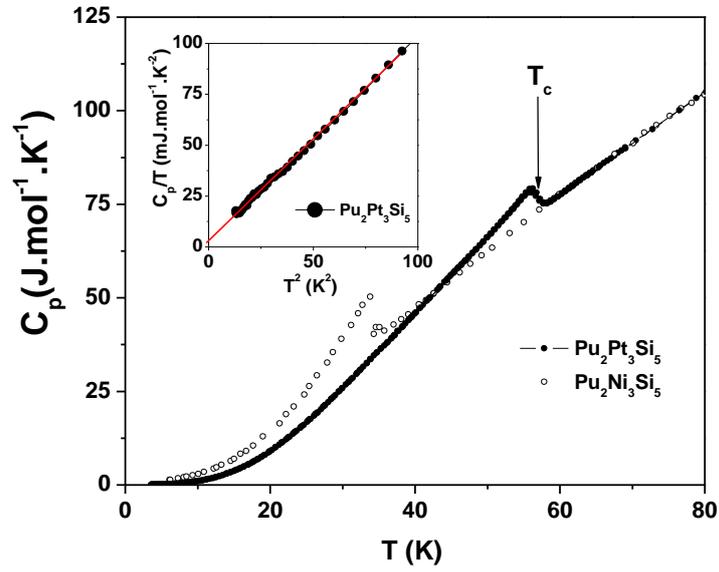

Figure 5 (Color online) Temperature dependence of the heat capacity of $Pu_2Pt_3Si_5$ and $Pu_2Ni_3Si_5$ showing a magnetic peak at $T_C$ and $T_N$ respectively. Data for $Pu_2Ni_3Si_5$ are extracted from Bauer et al. [29].
Inset: $C_p/T$ vs. $T^2$ for $Pu_2Pt_3Si_5$ at low temperature. The red line corresponds to the linear extrapolation, giving $\gamma_e \sim$ 2 mJ.mol$^{-1}$.K$^{-2}$/Pu.

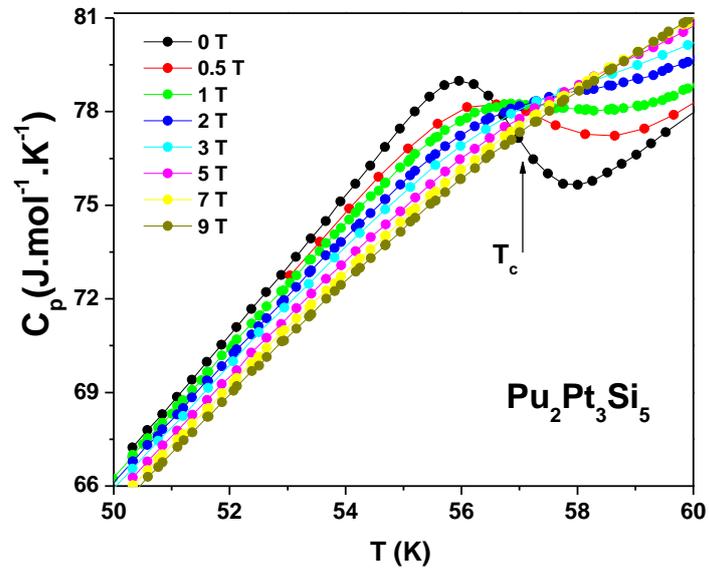

Figure 6 : (Color online) Magnetic field effect on the heat capacity $C_p(T)$ of $Pu_2Pt_3Si_5$ up to 9 T in the vicinity of the magnetic transition $T_C$. We observe a smearing out of the peak and a slight increase in position towards higher temperature.